\documentclass{svjour8}                    
\smartqed  
\usepackage{graphicx,amsmath,amssymb,amsbsy,latexsym,verbatim}
\newcommand{\be}{\nopagebreak[3]\begin{equation}}
\newcommand{\ee}{\end{equation}}
\newcommand{\ba}{\nopagebreak[3]\begin{eqnarray}}
\newcommand{\ea}{\end{eqnarray}}

\newcommand{\bc}{}

\begin{document}

\title{Atomism and Relationalism as guiding principles for Quantum Gravity}

\subtitle{Notes for the ``Seminar on the Philosophical Foundations of Quantum Gravity", Chicago, 27-29 September 2013 
}
\titlerunning{Discreteness and Relationalism} 

\author{Francesca Vidotto}
\institute{
Radboud University Nijmegen, Institute for Mathematics, Astrophysics and Particle Physics, 
Mailbox 79, P.O. Box 9010, 6500 GL Nijmegen, The Netherlands\\\email{fvidotto@science.ru.nl}
}
\thanks{The work of FV at Radboud University is supported
by a Rubicon fellowship from the Netherlands Organisation for Scientific Research (NWO).}

 \date{June 1, 2013}
\maketitle

\begin{abstract}
The research in quantum gravity has jauntily grown in the recent years, intersecting with conceptual and philosophical issues that have a long history. In this paper I analyze the conceptual basis on which Loop Quantum Gravity has grown, the way it deals with some classical problems of philosophy of science and the main methodological and philosophical assumptions on which it is based. In particular, I emphasize the importance that atomism (in the broadest sense) and relationalism have had in the construction of the theory.
\end{abstract}
\vskip3em

Scientists are guided in their investigation of Nature by ideas brought to them by philosophers. Not always scientists are aware of this, but still their way of working and the kind of questions that they address, have emerged in a humus of philosophical debates of long history.
Being aware of these debates, that shaped our present investigations, can help us to recognize paths and new directions for our science.

I think this is particularly true when the theory we are searching for requires a deep rethinking of basic concepts such as space and time. In order to analyze the recent  findings in quantum gravity, a theory that hopes to provide a new fundamental view on the nature of space and time, l start by a possible reconstruction of the thread from the ancient time to our modern debate.

\section{The historical framework: from Democritus to Einstein}

What does exist? Does space exists? or it emerges from the relations between \emph{bodies}
?
The idea that space can exist as a separate entity with respect to the bodies is not a primitive idea, but a revolution brought to humanity by Leucippus and Democritus in the V century BCE. They postulates the existence of elementary units, the \emph{atoms}, whose combination yields all the beautiful rainbow of different things that we observe in the world. The atoms randomly moves in a stage: this is space.  Space is here associated with the notion of vacuum. This is a contradictory notion, because gives an ontological status to the non-being. 
That's why this position was attached by Athens' school: Aristotle thought that Nature \emph{abhors a vacuum}. Plato did not like the atomistic/materialistic views either, at the point never to mention Democritus. Trough Greek decadence and the Christian era, only Atheniensis wisdom survived to the centuries.

The XVII-century debate on the nature of space should be framed in a culture dominated by the aristotelian/platonic thinking . The mainstream side of the debate was there taken by the \emph{relational} position, defended by Descartes and Leibniz, 
denying the existence of space but as a net of relations.
On the other side there was the substantivalist position proposed by Newton, in which Democritean bodies moves on an infinite immutable fixed empty space, according to the deterministic law of the new Mechanics. The empirical success of Mechanics has brought this position to became the mainstream one nowadays.

Nonetheless, Newton hesitated in proposing such an idea of space. He called this a working hypothesis (\emph{hypotheses non fingo}) and its enormous success as funding stone of the new developing science made the later scientists forget the doubts about it. What concerned  Newton more was that his law of gravitation, acting on an empty stage, were leading to the possibility of an \emph{action at distance}.
Only with the introduction of the notion of field this worry was removed: forces are fields, that permeate space.

Faraday and Maxwell described the electromagnetic phenomena as a manifestation of a field. The physics of the XX century took the notion of field to an ontological extreme: everything that exist is a manifestation of some field. So it is every particle, as discovered by Quantum Mechanics: a particle is just the excitation of a field, a manifestation of the quantum nature of every field. So it is space, as discovered by Einstein: space and time are the expression of the gravitational field.

 \begin{table}[b]
\centerline{\includegraphics[height=8cm]{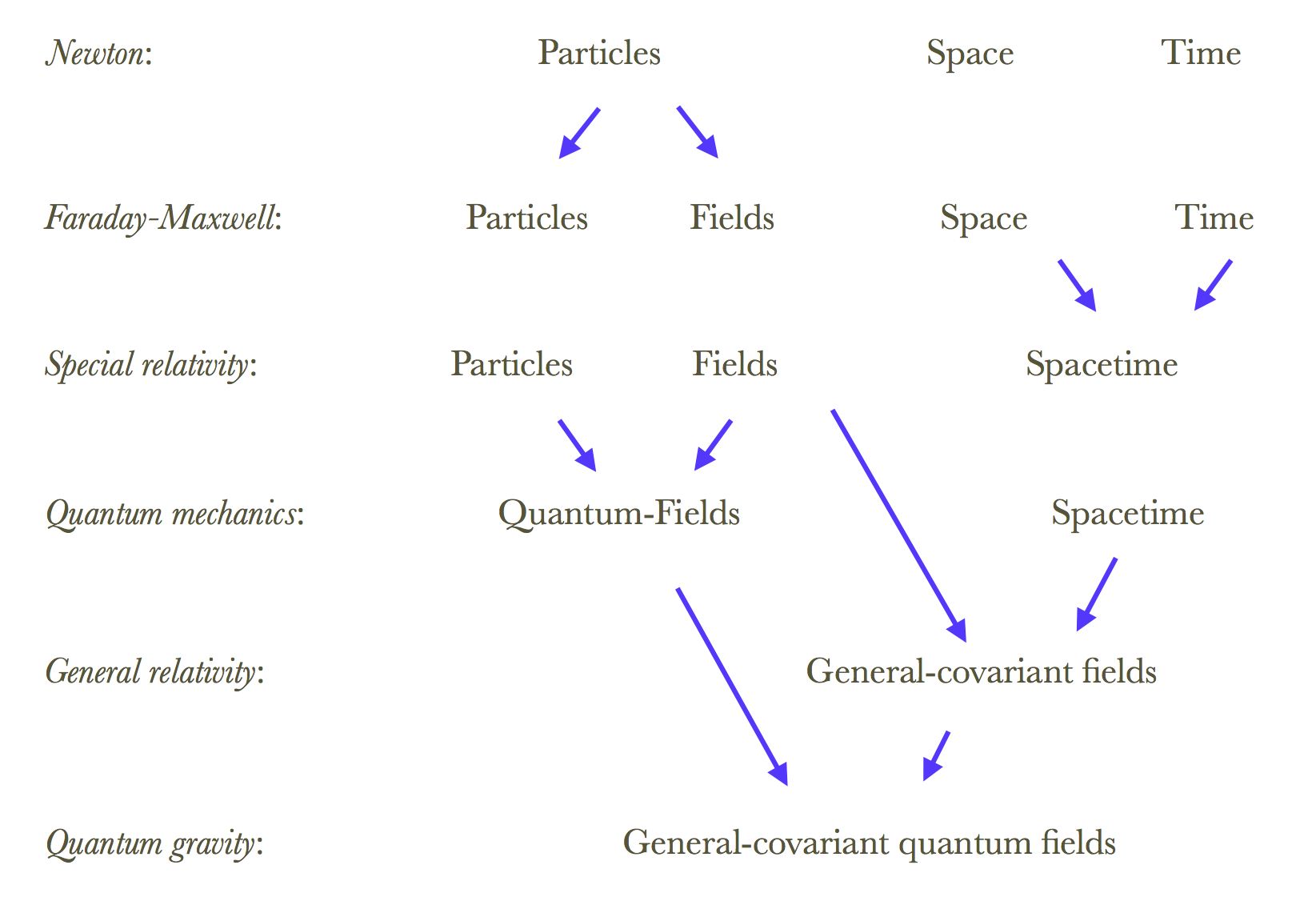}}
\end{table}%

General Relativity identifies spacetime and the gravitational field. This field, like all fields, should exhibit quantum properties at some scale, therefore space and time must have quantum properties as well.  This is the beauty and the difficulty of quantum gravity: it obliges us to a complete rethinking of what we mean by space and time. In order to do this, we need to sharp our description of quantum fields in order to make it covariant (i.e. compatible with General Relativity).
We have to learn a new language for describing the world. A language which is neither that of standard field theory on flat spacetime, nor that of Riemannian continuous geometry as space presents the discreteness typical of every quantum system.  We have to understand what is quantum space and what is quantum time.

\section{The end of space and time, the beginning of quantum gravity} 

As often happens in science, the contemporary questioning of the nature of
space started with a mistake. The problem was to extend Heisenberg's 
uncertainty relations to fields.  In 1931 Landau and Peierls  suggested that once applied to fields, the uncertainty relation would imply that no component of a field at a given spacetime point could be measured with arbitrary precision \cite{Landau:1931uq}.  The intuition was that an arbitrarily sharp spatiotemporal localization would have been in contradiction with the Heisenberg uncertainty relations.  Niels Bohr guessed immediately, and correctly, that this suggestion was wrong.  To prove it wrong, he embarked in a research program with Rosenfeld, that lead to a classic paper \cite{Bohr:1933uq} in which the two proved that in a quantum field theory the Heisenberg uncertainty relations \emph{do not} prevent a component of a field to be measured with arbitrary precision at a spacetime point. 
The Bohr-Rosenfeld analysis was done using the electromagnetic field: what if repeated with the gravitational field? This question engaged Landau's friend Bronstein \cite{Gorelik:1994fk} and he found that Landau's intuition in this case was correct \cite{Bronstein:9vn,Bronstein:1936kx}. This is the beginning and the core of quantum gravity.

In modern terms, Bronstein's argument would be the following. 
In order to measure some field value at a location $x$, its location should be determined with some precision $L$.  If this is done by having a particle at $x$, the quantum nature of the particle implies that there is an uncertainties $\Delta x$ and $\Delta p$ associated to its position and its momentum. To have the location determined with precision $L$, this should be greater than $\Delta x$, and since Heisenberg uncertainty gives $\Delta x >\hbar/\Delta p$, we have $\Delta p>\hbar/L$. The average absolute value of the momentum cannot be smaller than its fluctuation, therefore $|p| >\hbar/L$. This is a very well known consequence of Heisenberg uncertainty:  sharp location requires large momentum. In turn, large momentum implies large energy $E$. In the relativistic limit, where rest mass is negligible, $E\sim c p$. Sharp localization requires large energy.  

In General Relativity  any form of energy $E$ acts as a gravitational mass $M\sim E/c^2$ and distorts spacetime around itself. The distortion increases when energy is concentrated, to the point that a black hole forms when a mass $M$ is concentrated in a sphere of radius $R \sim G M/c^2$, where $G$ is the Newton constant. If $L$ arbitrary small in order to get a sharper localization, the concentrated energy will grow to the point where $R$ becomes larger than $L$. But in this case the region of size $L$ that we wanted to mark will be hidden beyond a black hole horizon, and we loose localization.  Therefore $L$ can be decreased only up to a minimum value, which clearly is reached when the horizon radius reaches $L$, that is when $R=L$. 

Combining the relations above, we obtain that the minimal size where we can localize a quantum particle without having it hidden by its own horizon, is 
\be
L=\frac{MG }{c^2}=\frac{EG }{c^4}=\frac{pG}{c^3}=\frac{\hbar G}{Lc^3}.
\ee
Solving this for $L$, we find that it is not possible to localize anything with a precision better than the length 
\be
l_o\ \sim\ \sqrt{\frac{\hbar G}{c^3}} \ \sim \ 10^{-33} \ cm,
\ee
which is called the Planck scale. Above this length scale, we can treat spacetime as a smooth space.  Below this scale, it makes no sense to talk about distance or extension.  

This simple derivation, using only semiclassical physics, characterizes the physics of quantum spacetime. 
The existence of a minimal length scale is the main feature of quantum gravity and gives it a universal character, analogous to Special Relativity and Quantum Mechanics.  Special Relativity can be seen as the discovery of the existence of a maximal local physical velocity, the speed of light $c$.  Quantum Mechanics can be interpreted as the discovery of a minimal action $\hbar$ in all physical interactions, or equivalently the fact that a finite region of phase space contains only a finite number of distinguishable (orthogonal) quantum states, and therefore there is a minimal amount of information in the state of a system. Quantum gravity is the discovery that there is a minimal length.

In Bronstein's words  \cite{Bronstein:9vn}: 
``Without a deep revision of classical notions it seems hardly possible to extend the quantum theory of gravity also to
[the short-distance] domain." Bronstein's result forces us to take seriously the connection between gravity and geometry. It shows that the Bohr-Rosenfeld argument,  showing that quantum fields can be defined in arbitrary small regions of space, fails in the presence of gravity.  Therefore the quantum gravitational field cannot be treated simply as a quantum field in space. The smooth metric geometry of physical space, which is the ground needed to define standard quantum field, is itself affected by quantum theory. What we need is a genuine quantum theory of geometry.

\section{Quanta of spacetime}

Background independence is at the core of General Relativity: spacetime 
is not a stage where fields interact, but it is an interacting field as the others. The modern understanding of fields is that they are associated to some gauge symmetry. In the Standard Model, %
these symmetries are respectively $U(1)$ for electromagnetic interactions, $SU(2)$ for weak interactions and $SU(3)$ for strong interactions. Gravitational interactions can also be characterized in terms of a gauge symmetry, as the others.

General Relativity is traditionally formulated in using the metric, but it admits an equivalent formulation where the fundamental object are reference fields: the tetrads. The metric can always be expressed in terms of the tetrads, therefore it is always possible to pass from the metric formulation to the tetrad one. On the other hand, only in the tetrad formulation fermions can be coupled to the gravitational field. A description of reality cannot be complete in a framework that does not includes fermions, so General Relativity in tetrad variables assumes somehow a more fundamental form.

So consider now the tetrad formulation. 
The invariance under diffeomorphisms implies the independence by coordinate transformations. This means that for each point of spacetime there is associated a tetrad that is locally Lorentz invariant. We can reduce ourselves to the rotational part of the Lorentz transformation, since time is pure gauge in General Relativity and we can always fix this gauge.
Is this gauge invariance, that naturally arises in the classical gravitational theory, the starting point for the quantum theory.
The quantum states have to be thought as boundary states \cite{Oeckl:2003vu,Oeckl:2005bv}, describing the geometry at some fixed time.
In the quantum theory, the tetrad turns out to be the generator of SU(2) transformations, satisfy the algebra of angular momentum. This implies %
that spacetime is quantized with a discrete spectrum. Notice that now we don't have any more a tetrad for each spacetime point, but a tetrad for each spacetime quanta. For each to these, the orientations of the reference fields is not relevant, only the relations between adjacent quanta matter. In Loop Quantum Gravity a spinnetwork state \cite{Rovelli:1995ac,Introduction} is state that is defined by the invariance under the rotations of the triads, the excitations of each spacetime quanta (its spin, corresponding to its physical size) and the adjacency relation between them (coded in an abstract graph).

As in every Yang-Mills theory, the gauge invariance of the triads brings in a gauge field. This is %
an object in the Lorentz algebra, %
that codes the information %
about intrinsic and extrinsic curvature. 
In the quantum theory the gauge invariant observables are defined by the path-ordered exponential of the gauge field. In the language of particle physics, this is a Wilson loops  (from this the name Loop for the quantum theory). In the language of differential geometry, it is called holonomy, namely the parallel transport of the gauge field  seen as a connection over the $SU(2)$ principal bundle. The canonical analysis of the theory shows that this is the conjugate variable to the triad. 

The variables used in Loop Quantum Gravity are group variables, as the variables used to describe the other fundamental interactions. The group is $SU(2)$, that is compact yielding a discrete spectrum for the corresponding observables. In particular, the geometry can be described trough observables such as areas, volumes and angles, constructed starting from the operator corresponding to the triads.

\section{Quantum relations}

We want now to discuss the dynamical properties of spacetime. 
In classical mechanics, a dynamical system can be defined
trough the relations between initial and final coordinates and momenta, defining the allowed trajectories.  In Quantum Mechanics, the trajectories between interactions are not deducible from the interaction outcomes, and the theory describes \emph{processes}.  The quantum analogs of the relations between values of physical variables at the \emph{boundaries} are the transition amplitudes $W$, which determine probabilities of alternative sets of outcomes. 

Quantum Mechanics describes how physical systems affect one another in the course of interactions \cite{Rovelli:1995fv}.  It computes the probabilities for the different possible effects of such interactions.  A common language for describing such processes is in terms of ``preparation" and ``measurement".  But this anthropomorphic language is misleading \cite{Rovelli:1995fv}.   What happens at the boundary of a process if simply a physical \emph{interaction} of the system with another completely generic physical system.  Notice that the structure of the theory is largely determined by the fact that this description is consistent with arbitrary displacements of what we decide to consider as the boundaries between processes. 

Let us now apply this perspective to field theory and in particular to gravity.  The central idea is now to consider a \emph{finite} portion of the trajectories of a system, where ``finite" means finite in time but also in space (this is called the ``boundary formalism" \cite{Oeckl:2003vu,Oeckl:2005bv}).  Thus, consider a finite bounded region $M$ of spacetime. For a field theory on a given fixed spacetime, this simply means to consider the evolution of the field in a spacial box, with given boundary values at the boundaries of the box.  Therefore the transition amplitudes will be functions of the field values on the initial spacelike surface, the final spacelike surface, but also on the ``sides": the timelike surface that bounds the box. In other words, the transition amplitudes $W$ are functions of the values of the field on the entire boundary of the spacetime region $M$.  Formally, $W$ can be expressed as the Feynman path integral of the field in $M$, with fixed values on the boundary $\Sigma=\partial M$. Obviously, $W$ will depend on these values of the fields as well as on the (spacetime) shape and geometry of $\Sigma$. For instance, on the time lapsed between the initial and final surfaces. 

Now let us take this same idea to gravity.  Then what we want is the transition amplitude $W$ that depend on the value of the gravitational field (as well as any other field which is present) on the boundary $\Sigma$ of a spacetime region. Formally, this will be given by the Feynman path integral in the internal region, at fixed boundary values of the gravitational (and other) fields $\Sigma$.  How do we now specify the shape, namely the geometry, of $\Sigma$?  Quantum gravity teaches us that we have not to do it. 

In fact, the gravitational field on the boundary of $\Sigma$ already specifies the shape of $\Sigma$, as it includes all relevant metric informations that can be gathered on the surface itself. Therefore $W$ will be a function of the boundary fields and nothing else. 

This happens as a consequence of a general property of parametrized systems: the temporal information is stored and mixed among the dynamical variables, instead as being signed out and separated from other variables as in unparametrized Newtonian mechanics.  In the general relativistic context, this holds for temporal as well as for spacial locations: $W$ will not be a function of space and time variables, but simply a function of the gravitational field on the boundary $\Sigma$ (up to diffeomorphisms of $\Sigma$), which contains the entire relevant geometrical information on the boundary. 

Therefore in quantum gravity the quantum dynamics will be captured by a transition amplitude $W$ which is a function of the (quantum) state of the field on a surface $\Sigma$.   Intuitively, $W$ is the ``sum over geometries" on a \emph{finite} bulk region bounded by $\Sigma$: this is called a spinfoam.  The explicit form of $W$ is one of the most important result in quantum gravity of the last years \cite{Engle:2007qf,Engle:2007uq,Engle:2007wy,Freidel:2007py,Kaminski:2009fm,Livine:2007hc}.

\section{The relational structure of quantum gravity}\label{relations}

The formal structure of Quantum Mechanics is relational, because Quantum Mechanics gives probability amplitudes for processes, and a process is what happens between interactions  \cite{Rovelli:1995fv}. Therefore Quantum Mechanics describes the universe in terms of the way systems affect one another.  States are descriptions of manners a system can affect another system. Quantum Mechanics is therefore based on \emph{relations} between systems, where the relation is instantiated by a physical interaction.  

The structure of General Relativity is also relational, because the localization of dynamical objects is not given with respect to a fixed background structure; instead, \emph{bodies} are only localized with respect to one another, where \emph{bodies} includes all dynamical objects, included the gravitational field.  The relevant relation that builds the spacetime structure is of course contiguity: the fact of being ``next to one another" in spacetime.  We can view a general relativistic theory as a dynamical patchwork of spacetime regions adjacent to one another at their boundaries. 

A fundamental ingredient in XX century physics is locality:  interaction are  local, namely they require spacetime contiguity. But the contrary is also true: the only way to ascertain that two objects are contiguous is by means of having them interact. Therefore locality reveals a fundamental structural analogy between the relations on which Quantum Mechanics is based and those on which spacetime is based.  Quantum gravity makes this connection completely explicit. A process is not in a spacetime region: a process \emph{is} a spacetime region.  A state is not somewhere in space: it \emph{is} the description of the way two processes interact, or two spacetime regions pass information to one another. Viceversa, a spacetime region \emph{is} a process: because it is actually like a Feynman sum of everything can happen between its boundaries: \vspace{2mm}

\begin{center}
  \begin{tabular}{@{} ccc @{}}
    Quantum Mechanics & ¥ & General Relativity \\ 
    \hline
\\ 
    Process & ¥ & Spacetime region \\ 
    ¥ & $\leftarrow$ Locality $\rightarrow$ & ¥ \\ 
    State & ¥ &\hspace{2em} Boundary, space region \hspace{2em} \\ \\
    \hline
  \end{tabular}
\end{center}
\vspace{2mm}

This structural identification is in fact much deeper.  As noticed, the most remarkable aspect of quantum theory is that the boundary between processes can be moved at wish.  Final total amplitudes are not affected by displacing the boundary between ``observed system" and ``observing system". The same is true for spacetime: boundaries are arbitrarily drawn in spacetime.  The physical theory is therefore a description of how arbitrary partitions of nature affect one another. because of locality and because of gravity, these partitions are at the same time subsystems split and partitions of spacetime. A spacetime is a process, a state is what happens at its boundary.

\section{Conclusion}
What does exist? Does space exists? or it emerges from the relations between \emph{bodies}?
Quantum gravity is the quest for a synthesis between Quantum Mechanics and General Relativity. But while doing this, quantum gravity would achieve a synthesis also between substantivalism and relationalism: space is a field, that come to existence only trough its interactions.
Space is constitute by atoms of space, defined trough their relations.

\vskip2cm

\begin{acknowledgements}
I acknowledge support of the Netherlands Organisation for ScientiÞc Research (NWO) under their Rubicon program.
\end{acknowledgements}

\newpage

\bibliographystyle{hunsrt} 
\bibliography{bib,phil,BiblioCarlo}
\end{document}